\documentclass[prl,superscriptaddress,amsfonts,twocolumn,showpacs,floatfix]{revtex4-1}
\usepackage{amsmath}
\usepackage{amsfonts}
\usepackage{amssymb}
\usepackage{bm}
\usepackage{tabularx}
\usepackage[dvips]{graphicx,color}

\begin{document}
%%%%% Title %%%%%
\title{
Universal dynamics of magnetic monopoles in two-dimensional kagom\'{e} ice
}
%%%%%%%%%%%%%%%%%

%%%%% Authors %%%%%
\author{Hiroshi Takatsu}
%\email{takatsu@scphys.kyoto-u.ac.jp}
\affiliation{Department of Physics, Tokyo Metropolitan University, Hachioji-shi, Tokyo 192-0397, Japan}
\affiliation{Department of Energy and Hydrocarbon Chemistry, Graduate School of Engineering, Kyoto University, Kyoto 615-8510, Japan}
%\affiliation{Engineering Education Research Center, Graduate School of Engineering, Kyoto University, Kyoto 615-8510, Japan}

\author{Kazuki Goto}
\affiliation{Department of Physics, Tokyo Metropolitan University, Hachioji-shi, Tokyo 192-0397, Japan}

\author{Hiromi Otsuka}
\affiliation{Department of Physics, Tokyo Metropolitan University, Hachioji-shi, Tokyo 192-0397, Japan}

\author{Taku J. Sato}
\affiliation{Institute of Multidisciplinary Research for Advanced Materials, Tohoku University, 2-1-1 Katahira, Sendai 980-8577, Japan}

\author{Jeffrey W. Lynn}
\affiliation{NIST Center for Neutron Research, National Institute of Standards and Technology, Gaithersburg, MD 20899-6102, U.S.A}

\author{Kazuyuki Matsubayashi}
\affiliation{Graduate School of Informatics and Engineering, The University of Electro-Communications}
\affiliation{Institute for Solid State Physics, University of Tokyo, Kashiwanoha 5-1-5, Kashiwa, Chiba 277-8581, Japan}

\author{Yoshiya Uwatoko}
\affiliation{Institute for Solid State Physics, University of Tokyo, Kashiwanoha 5-1-5, Kashiwa, Chiba 277-8581, Japan}

\author{Ryuji Higashinaka}
\affiliation{Department of Physics, Tokyo Metropolitan University, Hachioji-shi, Tokyo 192-0397, Japan}

\author{Kazuyuki Matsuhira}
\affiliation{Department of Electronics, Faculty of Engineering, Kyushu Institute of Technology, Kitakyushu 804-8550, Japan}

\author{Zenji Hiroi}
\affiliation{Institute for Solid State Physics, University of Tokyo, Kashiwanoha 5-1-5, Kashiwa, Chiba 277-8581, Japan}

\author{Hiroaki Kadowaki}
\affiliation{Department of Physics, Tokyo Metropolitan University, Hachioji-shi, Tokyo 192-0397, Japan}
%%%%%%%%%%%%%%%%%%%
\date{\today}

%%%%% Abstract %%%%%
\begin{abstract}
A magnetic monopole in spin ice is a novel quasiparticle excitation in condensed matter physics, and we found that the ac frequency dependent magnetic susceptibility $\chi(\omega)$ in the two-dimensional (2D) spin ice (so-called kagom\'{e} ice) of Dy$_2$Ti$_2$O$_7$ shows a single scaling form. 
This behavior can be understood in terms of the dynamical scaling law for 2D Coulomb gas (CG) systems [Phys. Rev. B ${\bm {90}}$, 144428 (2014)], characterized by the charge correlation length $\xi (\propto1/\sqrt{\omega_1})$, where $\omega_{1}$ is a characteristic frequency proportional to the peak position of the imaginary part of $\chi(\omega)$. It is a generic behavior among a wide variety of models such as the vortex dynamics of 2D superconductors, 2D superfluids, classical XY magnets, and dynamics of melting of Wigner crystals.
\end{abstract}
%%%%%%%%%%%%%%%%%%%%

\maketitle

\begin{figure*}[tbp]
\begin{center}
\includegraphics[width=0.90\textwidth]{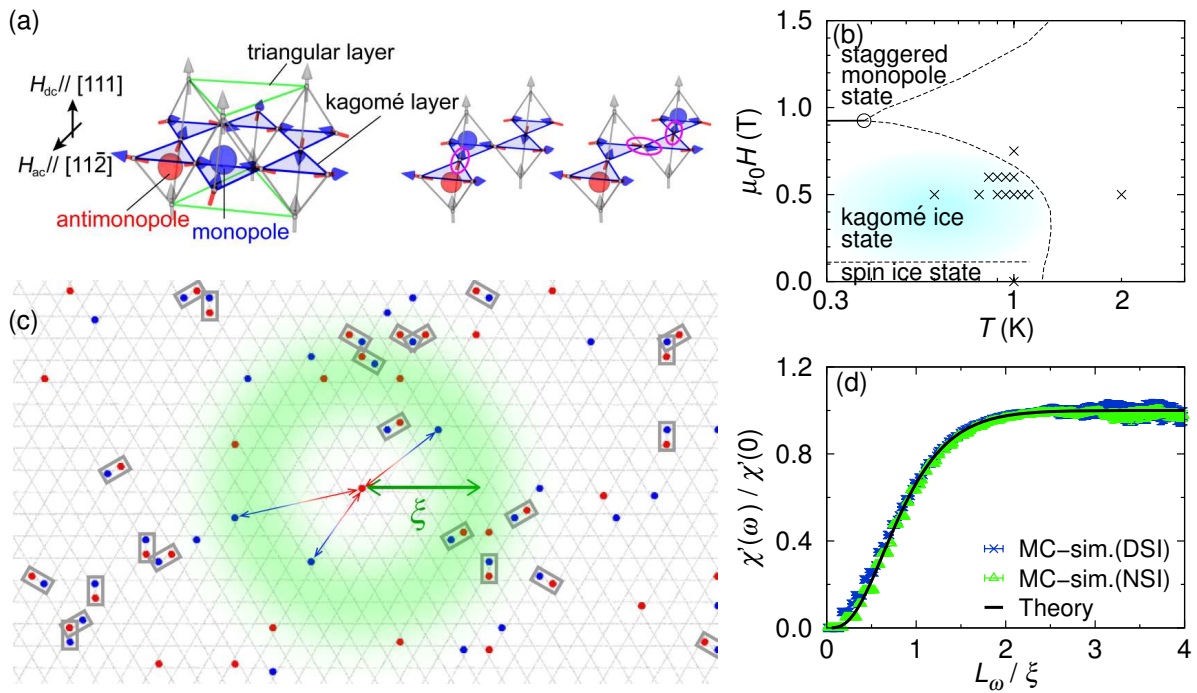}
\end{center}
\caption{
(a) Magnetic monopoles in kagom\'{e} ice. 
Left: a monopole and antimonopole pair is created in the KI state by 
a single spin flip. Right: they separate by successively flipping spins. 
(b) Phase diagram of DTO under an applied filed along the 111 direction~\cite{HiroiJPSJ2003,HigashinakaJPSJ2004}. 
The solid line represents the first-order phase transition to the critical point (open circle). 
The dashed lines are crossover regimes showing the borders of the KI state. 
The typical KI range is $\mu_0H_{\rm dc}\sim 0.5$ T and $T<1.2$ K. 
Cross symbols indicate
measurement points of $\protect\chi(\protect\omega)$ in the $(T,H)$ plane. 
(c) A snapshot of monopoles (blue) and antimonopoles (red) in the MC simulation of the DSI model for DTO at $T=1$ K and $\mu_0H_{\rm dc}=0.5$ T. 
This is shown for a 2D kagom\'{e} plane viewed from the 111 direction. 
There are fractionalized monopoles and antimonopoles, separated by a certain distance $\sim\protect\xi$.
A monopole-antimonopole pair closely bounded in an adjacent region is also seen (gray squares). 
$\xi$ determined by the scaling plot of Fig.~\ref{fig.1}(d) 
agrees with the visual distance between separate monopoles in the snapshot (green arrow), $\xi\sim 30$~\AA~\cite{Supplemental_DTO}.
(d) Calculated $\chi '(\omega)$ 
plotted as a function of $L_{\omega}/\xi (= (\omega_1/\omega)^{1/2})$ for the NSI and DSI models and comparison with the theoretical scaling curve of Ref.~\onlinecite{H.OtsukaPRB2014}, 
$\protect\chi ^{\prime }(\protect\omega )/\protect\chi ^{\prime }(0)=\mathcal{F}^{\prime }(L_{\protect\omega}/\protect\xi )$. 
}
\label{fig.1}
\end{figure*}

%%%%%%%%%%%%%%%%%%%%%%%%
%%%%% Introduction %%%%%
%%%%%%%%%%%%%%%%%%%%%%%%
%%%%%% [1] %%%%%%%
For a decade or more, magnetic monopoles in condensed matters have attracted much attention as tractable analogs of magnetic charged particles~\cite{ZhongScience2003,QiScience2009,CastelnovoNature2008}. 
An example is the elementary excitation~\cite{CastelnovoNature2008} in spin ice (SI)~\cite{BramwellScience2001}, where point defects~\cite{RyzhkinJETP2005} or monopoles emerge from its macroscopically degenerate ground states by breaking local ``2-in--2-out'' spin configurations (Fig.~\ref{fig.1}(a)). 
A pair of monopoles and antimonopoles is fractionalized into two independent quasiparticle defects by successively flipping spins (Fig.~\ref{fig.1}(a)) and are thought to move diffusively along Dirac strings. Although they are not true elementary particles, the concept of monopoles in SI is attractive and their static and dynamic properties have been investigated by several experiments~\cite{MorrisScience2009,KadowakiJPSJ2009,FennellScience2009,note_Monopole_in_SI}. 

%%%%%% [2] %%%%%%%
Dy$_2$Ti$_2$O$_7$ (DTO) is a pyrochlore magnet showing the SI ground state~\cite{RamirezNature1999,BramwellScience2001}. 
It is known that the space dimensionality of the SI state can be tuned by applying a magnetic field along the 111 direction of the crystallographic axis of this compound~\cite{MatsuhiraJPCM2002}.
Along the 111 direction, DTO consists of alternating stacks of the triangular and kagom\'{e} lattice layers of Dy$^{3+}$ (Fig.~\ref{fig.1}(a)). 
Under an appropriate field $\sim0.5$~T at low temperatures, the spins on the triangular lattice layers are fixed parallel to the 111 direction, while the spins on each kagom\'{e} lattice layer remain frustrated, keeping the 2-in--2-out spin configuration (Fig.~\ref{fig.1}(a)). This state is referred to as the kagom\'{e} ice (KI) state (Fig.~\ref{fig.1}(b))~\cite{MatsuhiraJPCM2002,UdagawaJPSJ2002,HiroiJPSJ2003,SakakibaraPRL2003,HigashinakaJPSJ2004,TabataPRL2006,TakatsuJPJS2013-1} where the monopoles are restricted to move in each two-dimensional (2D) kagom\'{e} plane~\cite{TakatsuJPJS2013-1}. 
It is thus considered that the 2D dynamics of monopoles can potentially be evaluated in the KI state. 

%%%%%% [3] %%%%%%%
Recently, it has been theoretically pointed out that monopoles in the KI state behave like a 2D Coulomb gas (CG) and obey a scaling law for the frequency-dependent magnetic susceptibility $\chi(\omega)$~\cite{H.OtsukaPRB2014}:  
\begin{equation}
\chi(\omega)/\chi(0) = {\cal F}(\omega/\omega_1) = {\cal F}(L_{\omega}/\xi), 
\label{eq.scaling}
\end{equation}
where $\omega_1 = {\cal D}/\xi^2$ is a characteristic frequency 
determined by the charge correlation length $\xi$ of the KI and 
the diffusion constant ${\cal D}$. 
The frequencies $\omega$ is related to 
a mean diffusion length $L_{\omega}=({\cal D}/\omega)^{1/2}$~\cite{H.OtsukaPRB2014}.
This scaling law is universal among 2D CG systems such as the vortex dynamics of 2D superconductors~\cite{BeasleyPRL1979,EpsteinPRL1981}, 2D superfluids~\cite{AmbegaokarPRL1978,BishopPRL1978}, classical XY magnets~\cite{KosterlitzJPC1973,KosterlitzJPC1974}, and the dynamics of melting of Wigner crystals~\cite{StrandburgRMP1988}.
It is a novel perspective for the study of the monopole dynamics in SI compounds. 
However, there has been no detailed experiments for clarifying such dynamics and the scaling relation of $\chi(\omega)$.

%%%%%% [4] %%%%%%%
In this study, we experimentally investigated the monopole dynamics in the KI state of DTO by measuring the frequency dependence of the ac magnetic susceptibility $\chi(\omega)$.
These enable us to capture a new aspect of monopoles in 2D and to interpret experimental data of $\chi(\omega)$ clearly by a comparison with a theoretical scaling relation for generic systems of 2D charged particles.
We also examined this interpretation by Monte Carlo (MC) simulations and neutron scattering experiments.

%%%%%%%%%%%%%%%%%%%%%%%%%%%%%%%%
%%%%% Kagome spin ice state %%%%
%%%%%%%%%%%%%%%%%%%%%%%%%%%%%%%%
%%%%% Phase diagram %%%%%
%%%%%%%%%%%%%%%%%%%%%%%%%%%%%%%%%%%%%%%%%%%%%%%%%%%%%%%%%%%%%%%%%%%%%%%

%%%%%%%%%%%%%%%%%%%%%%%%%%%%%%%%%%%%
%%%%%%%%%  Experiments %%%%%%%%%%%%%
%%%%%%%%%%%%%%%%%%%%%%%%%%%%%%%%%%%%
Single crystals of Dy$_{2}$Ti$_{2}$O$_{7}$ were grown by a floating zone method. 
%%%%%
For the measurements of $\chi(\omega)$,
we used a thin rectangular sample with a wide plane including the $111$ and $11\bar{2}$ directions.
The sample was approximately $3\times5\times0.2$~mm$^3$ and weighted 19.6~mg,
which ensures a negligible demagnetization effect.
%%%%%
$\chi(\omega)$ was measured by a mutual inductance method down to 0.6~K. 
An ac magnetic field $\bm{H}_{\mathrm{ac}}$ with the strength of 0.01 or 0.1~mT-rms and 
frequencies ranging from $\omega/2\pi=0.5$ to 2000~Hz was applied along the $11\bar{2}$ direction using a small handmade coil.
A dc magnetic field $\bm{H}_{\rm dc}$ along the $111$ direction was generated 
by a horizontal split-pair magnet.
%%%%%
The accuracy of the sample alignment with respect to  $\bm{H}_{\rm dc}$ and $\bm{H}_{\mathrm{ac}}$
is within 1$^\circ$. 
%%%%%
The absolute values of $\chi(\omega)$ were estimated by comparing the data obtained by 
a commercial SQUID magnetometer (Quantum design, MPMS).
%%%%%
%
%%%%%
Neutron scattering experiments on a single crystalline sample under $\bm{H}_{\rm dc}\parallel111$ were performed 
on the triple-axis spectrometer BT-7 (Ref.~\onlinecite{LynnJNIST2012}) at the NIST Center for Neutron Research.
The sample had a size of approximately $22\times3.1\times0.58$~mm$^3$
for a long direction parallel to the 111 direction, which ensured a negligible demagnetization effect.
It was mounted in a dilution refrigerator so as to measure the scattering plane perpendicular to 
the 111 direction.
Scans around a pinch point at $(8/3,-4/3,-4/3)$ were carried out using a position-sensitive detector.
The obtained data were corrected for background and absorption.
Uncertainties where indicated represent one standard deviation.
%%%%%
%
%%%%%
MC simulations were carried out on both nearest-neighbor spin ice (NSI) and dipolar spin ice (DSI) models
using the same method as that in previous studies~\cite{KadowakiJPSJ2009,TakatsuJPJS2013-1,TakatsuJPJS2013-2,TabataPRL2006,Supplemental_DTO}. 
For calculating $\chi(\omega)$,
we simulated a system of 36$\times$36 pairs of tetrahedra (15552 spins) for the NSI model and
48$\times$48 pairs of tetrahedra (27648 spins) for the DSI model using periodic boundary conditions:
the length on a side of these systems are up to 250--340~\AA,
which is large enough for $\xi\sim30$~\AA\, at $(T,H)= (1~\rm{K}, 0.5~\rm{T})$.
For calculating the intensity of neutron scattering,
we carried out MC simulations on a system size up to 110592 spins, using the DSI model.

%%%%%%%%%%%%%%%%%%%%%%%%%%%%%%%%%%%%%%%%%%%%%%%
%%%%%%%%%  Results and Discussion %%%%%%%%%%%%%
%%%%%%%%%%%%%%%%%%%%%%%%%%%%%%%%%%%%%%%%%%%%%%%
%%%%%%%%%%%%%%%%%%%%%%%%%%%%%%%%%%%%%%%%%%%%%%%%%%%%%%%%%%%%%%%%%%%%%
%%%%%%%%% Applicability of the theoretical scaling relation %%%%%%%%%
%%%%%%%%%%%%%%%%%%%%%%%%%%%%%%%%%%%%%%%%%%%%%%%%%%%%%%%%%%%%%%%%%%%%%
%%%%% [1] %%%%%
%
Let us first evaluate and discuss the applicable range of the theoretical scaling relation for actual SI compounds, since it is known that there are two models for SI (i.e., the NSI and DSI models) and the theoretical scaling relation is based on the NSI model~\cite{H.OtsukaPRB2014}.
The NSI model is the simplest SI model based on the Ising model only with the nearest-neighbor (NN) ferromagnetic exchange coupling on the pyrochlore lattice, by which the frustration of SI~\cite{BramwellScience2001} and KI~\cite{MatsuhiraJPCM2002} is most simply explained. On the other hand, DTO is well represented by the Ising model with a dipolar interaction (i.e., the DSI model)~\cite{HertogPRL2000,MelkoPRL2001}. 
Although DSI behaves almost identically to NSI in the ground-state manifold obeying the ice rule~\cite{IsakovPRL2005}, there are certain differences for ice-rule breaking monopoles.
Specifically, in the KI state for NSI the entropic 2D Coulomb potential~\cite{IsakovPRB2004} $\propto T \ln(r)$ works between two monopoles separated with a long distance $r$, while for DSI an additional three dimensional Coulomb potential~\cite{CastelnovoNature2008} $\propto -r^{-1}$ is dominant at short distances and at low temperatures. 
To illustrate this situation, in Fig.~\ref{fig.1}(c), we show a snapshot of a MC simulation of the DSI model of DTO at $(T,H)= (1~\rm{K}, 0.5~\rm{T})$. One can recognize two types of monopoles, i.e., monopole-pairs bound by the $-r^{-1}$ potential, and those separated by $\sim \xi$ exhibiting the diffusive motion in the $T \ln(r)$ potential. 
We thus expect that characteristics of the NSI and DSI under $\bm{H}_{\rm dc}\parallel111$ are similar at long distances, where the dynamical scaling theory of Ref.~\onlinecite{H.OtsukaPRB2014} can be applied to the DSI model and DTO.

%%%%% [2] %%%%%
%
To further quantify the expectation, we have calculated the real part of the magnetic susceptibility $\chi '(\omega)$ for the NSI and DSI models using MC simulations~\cite{Supplemental_DTO}. Figure~\ref{fig.1}(d) shows that $\chi '(\omega)$ of these two models reasonably agree with each other at long distances ($L_{\omega}/\xi > 0.5$). Moreover, these results are  compatible with the theoretical scaling curve (Eq.~(12) of Ref.~\onlinecite{H.OtsukaPRB2014}) at long distances. They agree remarkably well in $L_{\omega}/\xi > 1$ and reasonably well in $1>L_{\omega}/\xi > 0.5$.
It is thus considered that the theoretical scaling relation of Eq.~(\ref{eq.scaling}) (Eqs.~(12) and (14) of Ref.~\onlinecite{H.OtsukaPRB2014}) is applicable for the monopole dynamics at long distances (i.e., low frequencies, $L_{\omega} \propto 1/\sqrt{\omega}$) and not-very-low temperatures. We notice that there is such a range close to $(T,H)= (1~\rm{K}, 0.5~\rm{T})$ for DTO.

%%%%%%%%%%%%%%%%%%%%%%%%%%%%%%%%%%%%%%%%%%%%%%%%%%%%%%%%%%%%%%%%%%%%%%%%%%%%%%%
%%%%%%%%% Frequency dependent magnetic susceptibility in the KI state %%%%%%%%%
%%%%%%%%%%%%%%%%%%%%%%%%%%%%%%%%%%%%%%%%%%%%%%%%%%%%%%%%%%%%%%%%%%%%%%%%%%%%%%%
%
%%%%% [1] %%%%%
%%%%% Frequency dependence of ac susceptibility at 0.5 T for various temp. %%%%%
\begin{figure}[tbp]
\begin{center}
\includegraphics[width=0.45\textwidth]{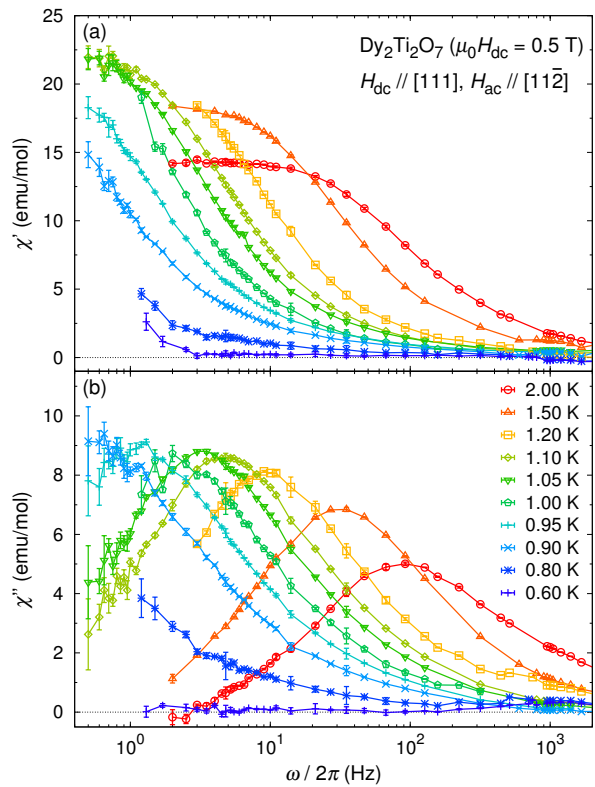}
\end{center}
\caption{ Frequency dependence of the magnetic susceptibility in the
kagome plane at various temperatures under $\mu_0H_{\rm dc} = 0.5$~T. 
(a) Real part $\protect\chi ^{\prime }(\protect\omega )$. 
(b) Imaginary part $\protect\chi ^{\prime \prime }(\protect\omega )$.
Note: 1 emu/(mol-Oe) = 4$\pi$ 10$^{-6}$ m$^3$/mol.}
\label{fig.2}
\end{figure}
%%%%%%%%%%%%%%%%%%%%%%%%%%%%%%%%%%%%%%%%%%%%%%%%%%%%%%%%%%%%%%%%%%%%%%%%%%%%%%%%%
Figure~\ref{fig.2} shows the frequency dependence of the real and imaginary
parts of the dynamic susceptibility $\chi (\omega )=\chi ^{\prime }(\omega)+i\chi ^{\prime \prime }(\omega )$ 
at various temperatures for $\mu_0H_{\rm dc}=0.5$~T along the 111 direction.
At this field, the system is in the KI state for $T<1.2$~K (Fig.~\ref{fig.1}(b)). As the temperature is lowered, the peak position  
of $\chi ^{\prime \prime }(\omega )$, 
which appears approximately at $\omega =1.4\omega_{1}$~\cite{H.OtsukaPRB2014}, shifts to lower values, i.e., the spin dynamics is slowed down. 

\begin{figure}[tbp]
\begin{center}
\includegraphics[width=0.45\textwidth]{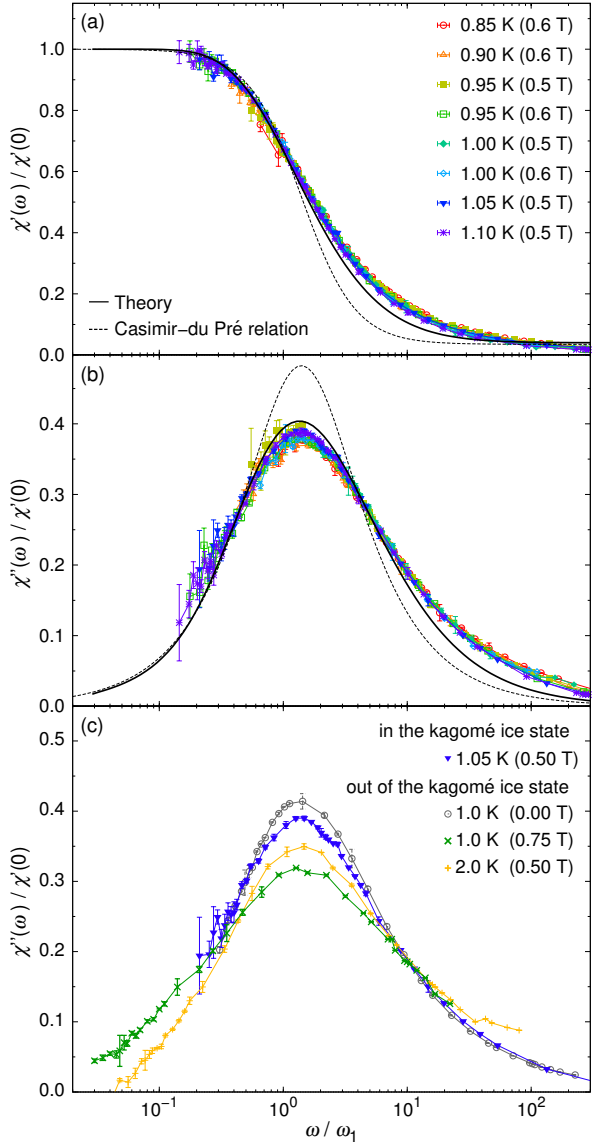}
\end{center}
\caption{ 
The frequency dependence of the magnetic susceptibility. 
(a) $\protect\chi^{\prime }(\protect\omega )$ and 
(b) $\protect\chi ^{\prime \prime }(\protect%
\omega )$ inside the crossover borders of the KI state 
(Fig.~\ref{fig.1}(b)). 
The solid lines are the theoretical scaling curves based on the
2D CG model\protect\cite{H.OtsukaPRB2014}. The dashed lines represent the
simplest Casimir-du Pr\'{e} relaxation relation\protect\cite{Casimir1938}. 
(c) Comparison of the scaled $\protect\chi^{\prime \prime}(\protect%
\omega )$ inside the KI state with its deviation outside the KI state.}
\label{fig.3}
\end{figure}

%%%%% [2] %%%%%
We performed a scaling analysis of $\chi ^{\prime }(\omega )$ and 
$\chi^{\prime \prime }(\omega )$; the resulting scaling plots are shown in 
Figs.~\ref{fig.3}(a), \ref{fig.3}(b) and \ref{fig.3}(c) for the data inside and outside the
crossover regimes of the KI state (defined in Fig.~\ref{fig.1}(b)). 
The vertical and horizontal axes of these plots are normalized by 
$\chi ^{\prime}(\omega \rightarrow 0)$~\cite{Supplemental_DTO} for the dc limit and $\omega _{1}$, respectively.
One can clearly see that $\chi ^{\prime }(\omega /\omega _{1})/\chi ^{\prime
}(0)$ and $\chi ^{\prime \prime }(\omega /\omega _{1})/\chi ^{\prime }(0)$
inside the KI state region collapse onto a single scaling curve. On the
other hand, $\chi ^{\prime \prime }(\omega /\omega _{1})/\chi ^{\prime }(0)$
outside the KI phase region shows obvious deviations from scaling. These
results clearly indicate that the monopole dynamics is governed by a single
scaling function in the KI phase. %
%
%%%%% Scaling plot of the ac susceptibility %%%%%
%%%%%%%%%%%%%%%%%%%%%%%%%%%%%%%%%%%%%%%%%%%%%%%%%%%%%%%%%%%%%%%%%%%%%%%

%%%%% [3] %%%%%
In Figs.~\ref{fig.3}(a) and \ref{fig.3}(b), 
we show the theoretical scaling
curves for the 2D CG model~\cite{H.OtsukaPRB2014} to
compare with the experimental results. 
A very small constant term ($\sim0.04\chi^{\prime}(0)$~\cite{Supplemental_DTO}) was added to the theoretical $\chi^{\prime}(\omega)$ (Eq.(7) of Ref.~[\onlinecite{H.OtsukaPRB2014}]) to obtain a better fit to the experimental data (Fig.~\ref{fig.3}a).
This could be reasonable, because we have to take account of the fact that the
theoretical curve is the leading $B_{1}$-breather term of the form factor
perturbation~\cite{H.OtsukaPRB2014} and thus is expected to be a good approximation at low
frequencies $\omega <\omega _{1}$ or at long distances $L_{\omega }>\xi $
(Fig.~\ref{fig.1}(d)). In addition, there is another contribution from
NN bound-pair monopoles for DSI (Fig.~\ref{fig.1}(c)), which gives a small constant term to
 $\chi ^{\prime }(\omega )$ in the present low-frequency range.
We also show fits to the simplest Casimir-du Pr\'{e} relaxation
relation~\cite{Casimir1938}, 
$\chi (\omega )=\chi_{\mathrm{S}}+(\chi_{\mathrm{T}}-\chi_{\mathrm{S}})/(1+i\omega \tau ),$ 
where $\chi_{\mathrm{S}} $ and $\chi_{\mathrm{T}}$ are the adiabatic and isothermal
susceptibilities, respectively. This relation is thought to provide a good
description for the dielectric constant of gases and electrical charges with
mechanical oscillation with a single relaxation time $\tau$, but fails to reproduce the present experimental
results. 
Therefore, we can conclude that the observed scaling behaviors of $\chi^{\prime }(\omega )$ and $\chi ^{\prime \prime }(\omega)$ clearly
demonstrate that the dynamical response of monopoles in the KI phase of DTO
is characterized by single $\omega_1$ and 
equivalent to that of the 2D CG~\cite{H.OtsukaPRB2014}.

%%%%%%%%%%%%%%%%%%%%%%%%%%%%%%%%%%%%%%%%%%
%%%%% Neutron scattering experiments %%%%%
%%%%%%%%%%%%%%%%%%%%%%%%%%%%%%%%%%%%%%%%%%
%
%%%%% [1] %%%%%
%%%%% Neutron scattering %%%%%
\begin{figure}
\begin{center}
\includegraphics[width=0.45\textwidth]{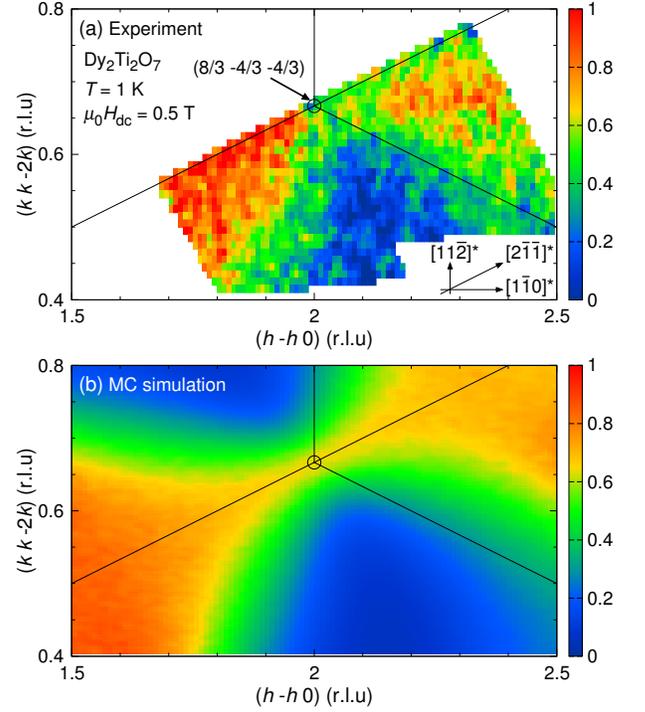}
\caption{
Neutron scattering intensity pattern.
(a) Intensity pattern close to a pinch point at (8/3,-4/3,-4/3) measured at $T=1$~K and $\mu_0H_{\rm dc} = 0.5$~T.
(b) Intensity pattern calculated by the MC simulation at the same temperature and field as those in (a).
}
\label{fig.4}
\end{center}
\end{figure}
%%%%%%%%%%%%%%%%%%%%%%%%%%%%%%%%%%%%%%%%%%%%%%%%%%%%%%%%%%%%%%%%%%%%%%%
%
In order to confirm that the physical picture of 2D monopoles, shown by the experimental scaling result of $\chi(\omega)$, is in the KI regime,
neutron scattering experiments were also conducted under $\bm{H}_{\rm dc} \parallel 111$.
We measured the magnetic diffuse (energy integrated) scattering close to a pinch point at $(8/3,-4/3,-4/3)$, where the effects of the monopole scattering are expected to be large~\cite{KadowakiJPSJ2009}, in the scattering plane perpendicular to the 111 direction at $(T,H)= (1~\rm{K}, 0.5~\rm{T})$. 

%%%%% [2] %%%%%
%
Figures~\ref{fig.4}(a) and \ref{fig.4}(b) show the diffuse scattering intensity map, together with the calculated diffuse scattering map obtained using the MC simulation. 
These intensity maps are in reasonable agreement; in particular, the pinch-point singularity, which is the hallmark of the Coulomb phase~\cite{HenleyARCP2010} ($T\rightarrow 0$), can be seen in both maps. 
One interesting point, found from a detailed inspection of the maps, is that the pinch-point singularity exhibits a slight broadening, implying existence of a small number of unbound monopoles.
The circumstance shown in the snapshot of Fig.~\ref{fig.1}(c) is equivalent to the map data around the pinch point singularity of Fig.~\ref{fig.4}(b), since these data are calculated under the same conditions.
Therefore, the reasonably good agreement between the magnetic diffuse scattering maps of experiments and MC simulations suggests that the image like the snapshot of Fig.1(c) is realized in experiments, where the 2D dynamics of  monopoles separated by distances $> \xi$ ($\sim 30$~\AA) plays an important role for the scaling relation of $\chi(\omega)$.

%%%%%%%%%%%%%%%%%%%%%%%%%%%%%%%%%%%%%%%%%%%%%%%%%%%%%%%%%%%%%%%%%%%%%%%%%%%%%%%%%
%%%%% Monopole dynamics in the KI state and its relation to the 2D CG model %%%%%
%%%%%%%%%%%%%%%%%%%%%%%%%%%%%%%%%%%%%%%%%%%%%%%%%%%%%%%%%%%%%%%%%%%%%%%%%%%%%%%%%
%%%%% [1] %%%%%
%
For decades, experimental investigation of the dynamics of the 2D CG systems
has been limited to the temperature dependence of the vortex dynamics in
superfluid films~\cite{MinnhagenRMP1987}. The importance of the
frequency-dependent universal scaling of $\chi(\omega)$ has been noticed
by one of the present authors (HO)~\cite{H.OtsukaPRB2014}, which was
triggered by early-stage experimental data of KI suggesting the scaling
behavior. 
A fortuitous aspect of measuring $\chi (\omega)$ in KI of DTO is that the ac magnetic field in the kagom\'{e} plane perturbs the system as a simple driving force of the magnetic monopoles, and the monopoles really respond to the ac magnetic field as expected for the theoretical 2D CG model, in which charged particles interact via the logarithmic Coulomb potential. 
This potential originates from the entropic interaction in KI~\cite{MoessnerPRB2003,IsakovPRB2004}.
It is also advantageous that the frequency dependence of the magnetic response can be readily measured using standard low temperature experimental techniques.
To the best of our knowledge, the present finding of the scaling behavior of 
$\chi (\omega)$ is the first experimental demonstration of 
the universal dynamics of the 2D CG model, exemplified by the well-controlled magnetic model of KI.
%

%%%%% [2] %%%%%
%
A recent detailed study of DTO in zero-field showed a tendency towards 
long range order below $\sim0.5$~K~\cite{PomaranskiNaturePhys2013}. 
Although similar behavior could
happen for KI of DTO at lower temperatures, we think that the results of the
present experiments occurring at relatively high temperatures $\sim 1$~K do
not change by a potential lifting of the degeneracy of KI as $T\rightarrow 0$.
The degeneracy, or at least quasi-degeneracy, of KI of DTO is very
important in the sense that it brings about the entropic 2D Coulomb
interaction for monopoles in the KI state, 
which is dominant at distances $r\geq \xi$
and governs the dynamics of the 2D CG system.

%%%%%%%%%%%%%%%%%%%%%%
%%%%% Conclusion %%%%%
%%%%%%%%%%%%%%%%%%%%%%
In summary, we studied the ac-frequency dependent magnetic susceptibility $\chi(\omega)$ for one of the typical spin-ice compounds Dy$_2$Ti$_2$O$_7$ and found the universal scaling behavior of $\chi(\omega)$ under the dc magnetic field parallel to the 111 direction. 
This behavior is characteristic of the 2D motion of monopoles and the charge correlation length $\xi (\propto 1/\sqrt{\omega}_1)$; i.e., it originates from the length dependent property of monopole dynamics in 2D. 
The present result is the first experimental demonstration of the monopole dynamics in 2D that obeys the dynamical scaling law for a generic behavior of the 2D Coulomb gas systems. 
%%%%%%%%%%%%%%%%%%%%%%%%%%%
%%%%% Acknowledgement %%%%%
%%%%%%%%%%%%%%%%%%%%%%%%%%%
\section*{Acknowledgment}
\begin{acknowledgments}
The susceptibility measurements were performed using facilities of 
the Institute for Solid State Physics (ISSP), the University of Tokyo. 
%
%Work on BT7 was supported by the US-Japan Cooperative Program on Neutron Scattering.
This work was supported by JSPS KAKENHI Grant Numbers 22014010, 25400345, 2600399, 26400336, 26400399, 17H04849.
This work was also supported in part by the US-Japan Cooperative Program on Neutron Scattering.
The identification of any commercial product or trade name does not imply endorsement or recommendation by the National Institute of Standards and Technology.
\end{acknowledgments}

%\bibliography{reference_DTO}
\bibliography{20210322_DTO_AC_with_SM.bbl}

%%%%%%%%%% Merge with supplemental materials %%%%%%%%%%
\pagebreak
\widetext
\begin{center}
\textbf{\large Supplemental Material for ``Universal dynamics of magnetic monopoles in two-dimensional kagom\'{e} ice''}
\end{center}
%%%%%%%%%% Merge with supplemental materials %%%%%%%%%%
%%%%%%%%%% Prefix a "S" to all equations, figures, tables and reset the counter %%%%%%%%%%
\setcounter{equation}{0}
\setcounter{figure}{0}
\setcounter{table}{0}
\setcounter{page}{1}
\makeatletter
\renewcommand{\theequation}{S\arabic{equation}}
\renewcommand{\thefigure}{S\arabic{figure}}
\renewcommand{\bibnumfmt}[1]{[S#1]}
\renewcommand{\citenumfont}[1]{S#1}
%%%%%%%%%% Prefix a "S" to all equations, figures, tables and reset the counter %%%%%%%%%%

\section{Abstract}
In this supplemental material, we describe details of Monte Carlo (MC) calculations of the ac magnetic susceptibility $\chi(\omega)$, and demonstrate the scaling relation of these MC results together with the theoretical curve of Ref.~[\onlinecite{H.OtsukaPRB2014_SM}], highlighting the applicable range of the dynamical theory to an actual spin ice material.
We also describe the estimation of the dc limit of $\chi '(\omega)$ and the slight correction of the theoretical curve to better fit to the experimental behavior of $\chi '(\omega)$.

%%%%% [2] %%%%%
\section{MC calculations of $\chi(\omega)$}
In MC calculations of $\chi(\omega)$, 
we used the following model dipolar Hamiltonian~\cite{HertogPRL2000_SM,MelkoJPCM2004_SM,RuffPRL2005_SM},
\begin{align}
\mathcal{H} = &-\mu_{\mathrm{eff}}\sum_{i,\alpha}\bm{S}_i^\alpha\cdot\bm{H} 
               - \sum_{\langle(i,\alpha),(i,\beta)\rangle}J_{i,\alpha;j,\beta} \bm{S}_i^\alpha\cdot\bm{S}_j^\beta \notag \\
              &+Dr_{nn}^3\sum_{i,j,\alpha,\beta}\biggl[\frac{\bm{S}_i^\alpha\bm{S}_j^\beta}{|\bm{R}_{ij}^{\alpha\beta}|^3}-\frac{3(\bm{S}_i^\alpha\cdot \bm{R}_{ij}^{\alpha\beta})(\bm{S}_j^b\cdot \bm{R}_{ij}^{\alpha\beta})}{|\bm{R}_{ij}^{\alpha\beta}|^5} \biggr],
\label{eq.S1}
\end{align}
where $\bm{S}_i^\alpha$, with unit length $|\bm{S}_i^\alpha| = 1$, 
represents the spin vector parallel to the local $\langle111\rangle$ direction
at the sublattice site $\alpha$ in the unit cell of the fcc lattice site $i$.
The first term represents the Zeeman interaction between the spins with 
the effective moment $\mu_{\mathrm{eff}}=9.866\mu_{\mathrm{B}}$ and the magnetic field $\bm{H}$.
The second and third terms are the exchange and dipolar interactions, respectively.

For the calculations of the dipolar spin ice (DSI) model,
we used the nearest-, second-, and third-neighbor interactions of
$J_{1} = -3.41$~K, $J_{2} = 0.14$~K, and $J_{3} = -0.025$~K, respectively,
and the dipolar interaction parameter $D=1.32$~K~\cite{YavorsPRL2008_SM,TabataPRL2006_SM,TakatsuJPJS2013-2_SM}.
For the calculations of the nearest-neighbor spin ice (NSI) model,
only one exchange interaction of $J_1=4.4$~K and $D=0$ were used (i.e., $J_2=J_3=0$)~\cite{HarrisPRL1997_SM,TabataPRL2006_SM}.
The best temperature and field regions were chosen as 
$\mu_0H_{\rm dc} = 0.3$~T and $T=0.6$ -- $0.8$~K for the NSI model,
and as 
$\mu_0H_{\rm dc} = 0.5$~T and $T=0.8$ -- $1.1$~K for the DSI model,
respectively, where the systems come into the kagom\'{e} ice state~\cite{KadowakiJPSJ2009_SM,TabataPRL2006_SM}.
In these MC calculations, 
a standard Metropolis algorithm with single-spin-flip dynamics was used~\cite{HertogPRL2000_SM,MelkoJPCM2004_SM,RuffPRL2005_SM}.

For calculating $\chi'(\omega)$,
we simulated a system of 36$\times$36 pairs of tetrahedra (15552 spins) for the NSI model and
48$\times$48 pairs of tetrahedra (27648 spins) for the DSI model using periodic boundary conditions:
to satisfy the theoretical assumption that $\bm{M}\simeq\bm{P}$~\cite{H.OtsukaPRB2014_SM},
we employed the systems with the length on a side up to 250--340~\AA\,,
large enough for 
$\xi\sim55$~\AA\, at $(T,H)= (0.8~\rm{K}, 0.5~\rm{T})$ or
$\xi\sim30$~\AA\, at $(T,H)= (1~\rm{K}, 0.5~\rm{T})$, determined by the scaling plots of $\chi'(\omega)$ and visual distance between separate monopoles in the snapshots of MC calculations under the DSI model.
Here, $\bm{M}$ is the uniform magnetization and
$\bm{P}$ is the total polarizability of magnetic monopoles.
From the dumbbell model~\cite{CastelnovoNature2008_SM},
$\displaystyle \bm{P} = q_{\rm D}\sum_{l}q_{l}\bm{x}_{l}$,
where $q_{l}$ is charge quantity [$q_{l}=+2$ (monopole) and $q_{l}=-2$ (antimonopole)], 
$\bm{x}_{l}$ is a site position of a monopole, $q_{\rm D}$ is the magnetic charge in the end of a dumbbell, 
$q_{\rm D} = 4\mu_{\rm eff}/\sqrt{3}a$, and $a$ is the lattice constant, respectively.
To calculate $\chi'(\omega)$ or $\chi'(L_\omega)$, we employed the same method as that in Ref.~[\onlinecite{H.OtsukaPRB2014_SM}]. 
Here, $L_{\omega}$ is the mean diffusion length~\cite{H.OtsukaPRB2014_SM},
which can be interpreted and treated as the distance that a pair of monopole and antimonopole is separated in a cycle of the ac magnetic field~\cite{H.OtsukaPRB2014_SM}. 
It is considered that these quasiparticle defects  move diffusively along Dirac strings~\cite{CastelnovoNature2008_SM,CastelnovoARCMP2012_SM}.

%%%%% [3] %%%%%
%%%%% Frequency dependence of ac susceptibility at 0.5 T for various temp. %%%%%
\begin{figure*}[tbp]
\begin{center}
\includegraphics[width=0.95\textwidth]{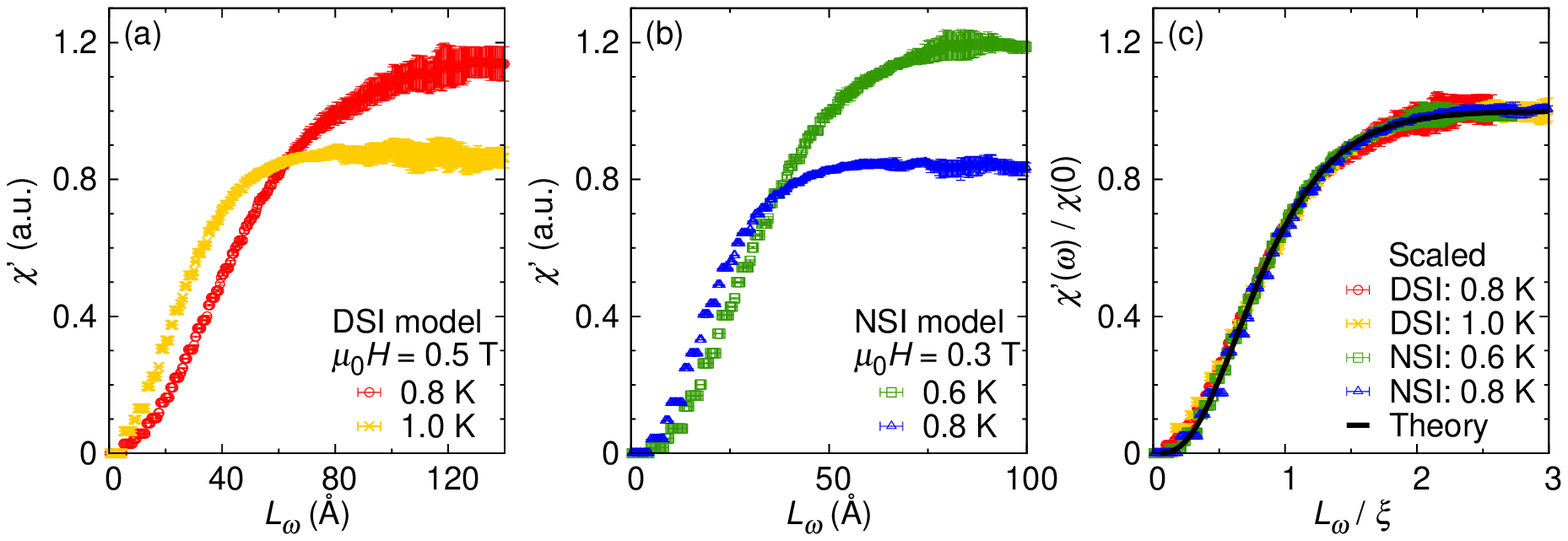}
\caption{
(a) Calculated results of the real part of the ac magnetic susceptibility $\chi'(\omega)$ 
based on the DSI model at $T=0.8$ and 1.0~K in $\mu_0 H = 0.5$~T.
(b) Calculated results of $\chi'(\omega)$ based on the NSI model $T=0.6$ and 0.8~K in $\mu_0 H = 0.3$~T.
(c) Scaled behavior of $\chi'(\omega)$ for both results of the DSI and NSI models, together with the theoretical curve~\cite{H.OtsukaPRB2014_SM}.
}
\label{fig.S1}
\end{center}
\end{figure*}
%%%%%%%%%%%%%%%%%%%%%%%%%%%%%%%%%%%%%%%%%%%%%%%%%%%%%%%%%%%%%%%%%%%%%%%%%%%%%%%%%
%
Results of the calculated $\chi'(\omega)$ on the basis of Eq.~(7) or Eq.(B4) in Ref.~[\onlinecite{H.OtsukaPRB2014_SM}]
are selected and shown in Figs.~S1(a) and S1(b),
plotted as a function of $L_{\omega}$,
for the DSI and NSI models.
In Fig.~S1(c), the scaling behavior of these curves is shown together with
the analytical description of the theoretical scaling curve (i.e., Eq.~(12) in Ref.~[\onlinecite{H.OtsukaPRB2014_SM}]).
By fitting this scaling relation to the calculated data with the adjustable parameter $\xi$,
we obtained $\xi = 32$~\AA\, for the DSI model at $(T,H)= (1.0, 0.5)$, for instance, which agrees with the visual distance between monopoles and antimonopoles in the snapshot (Fig.1(c) of the main text).
We confirmed that results of the MC calculations of $\chi'(\omega)$ are in quite good agreement with the theoretical curve,
particularly at $L_{\omega}/\xi>1$, the region of which is long-ranged distance.
A small discrepancy between the DSI model and the NSI model (and then the theoretical curve) is 
slightly visible at short distances below $L_{\omega}/\omega\simeq0.5$, as is discussed in the main text.
From these results,
we can conclude that the theoretical model of Ref.~[\onlinecite{H.OtsukaPRB2014_SM}] accounting for the 2D motion of monopoles
is applicable to the frequency-dependent dynamics of the DSI model and then to 
that of the actual spin-ice compound Dy$_2$Ti$_2$O$_7$ at long-ranged distances.

\section{Correction to theoretical scaling curve}
To better fit the experimental susceptibility $\chi^{\prime}_{\rm obs}(\omega)$ to the theoretical curve of Ref.~[\onlinecite{H.OtsukaPRB2014_SM}], a constant term $c_1$ (like an adiabatic susceptibility $\chi_{\mathrm{S}}$~\cite{Casimir1938_SM,ToppingJPCM2019_SM}) was needed in addition to the theoretical susceptibility $\chi^{\prime}_{\rm cal}(\omega)$ (Eq.(7) of Ref.~[\onlinecite{H.OtsukaPRB2014_SM}]), as rewritten below: 
\begin{align}
\chi^{\prime}_{\rm obs}(\omega) & = \chi^{\prime}_{\rm cal}(\omega) + c_1  \notag \\
                                & = \frac{\chi_{\rm c}}{T}\int_0^{L_{\omega}/4} r^{3}C(r)\,dr + c_1,
\label{eq.S2}
\end{align}
where  $\chi_{\rm c}$ is a constant, $L_{\omega}$ is a mean diffusion length, and 
$C(r)$ is a charge correlation function of monopoles~\cite{H.OtsukaPRB2014_SM}.
This correction could be reasonable as discussed in the main text.
Otherwise, it was necessary to subtract a few constant value $c_1$ 
from $\chi^{\prime}_{\rm obs}(\omega)$
and compare it with the theoretical scaling curve: i.e., 
$ (\chi^{\prime}_{\rm obs}(\omega) - c_1)/(\chi^{\prime}_{\rm obs}(0) - c_1) = \chi^{\prime}_{\rm cal}(\omega)/\chi^{\prime}_{\rm cal}(0)$. 
The following relations can thus be used for the fitting to experimental results:
\begin{align}
&\frac{\chi^{\prime}_{\rm obs}(\omega)}{\chi^{\prime}_{\rm obs}(0)} 
= \biggl(1- \frac{c_1}{\chi^{\prime}_{\rm obs}(0)} \biggr)\frac{\chi^{\prime}_{\rm cal}(\omega)}{\chi^{\prime}_{\rm cal}(0)}
+\frac{c_1}{\chi^{\prime}_{\rm obs}(0)}\\
&\frac{\chi^{\prime\prime}_{\rm obs}(\omega)}{\chi^{\prime}_{\rm obs}(0)} 
= \biggl(1- \frac{c_1}{\chi^{\prime}_{\rm obs}(0)} \biggr)\frac{\chi^{\prime\prime}_{\rm cal}(\omega)}{\chi^{\prime}_{\rm cal}(0)}.
\label{eq.S3}
\end{align}
In Fig.~3 of the main text, 
we show these corrected theoretical corves by Eqs.~(S3) and (S4).
Here, $c_1/\chi^{\prime}_{\rm obs}(0) = 0.04$ was used. 
Eqs.~(12) and (14) of Ref.~[\onlinecite{H.OtsukaPRB2014_SM}] were used for
$\chi^{\prime}_{\rm cal}(\omega)/\chi^{\prime}_{\rm cal}(0)$ 
and 
$\chi^{\prime\prime}_{\rm cal}(\omega)/\chi^{\prime}_{\rm cal}(0)$,
respectively.

\section{DC limit of the susceptibility}
%%%%% Frequency dependence of ac susceptibility at 0.5 T for various temp. %%%%%
\begin{figure}[t]
\begin{center}
\includegraphics[width=0.45\textwidth]{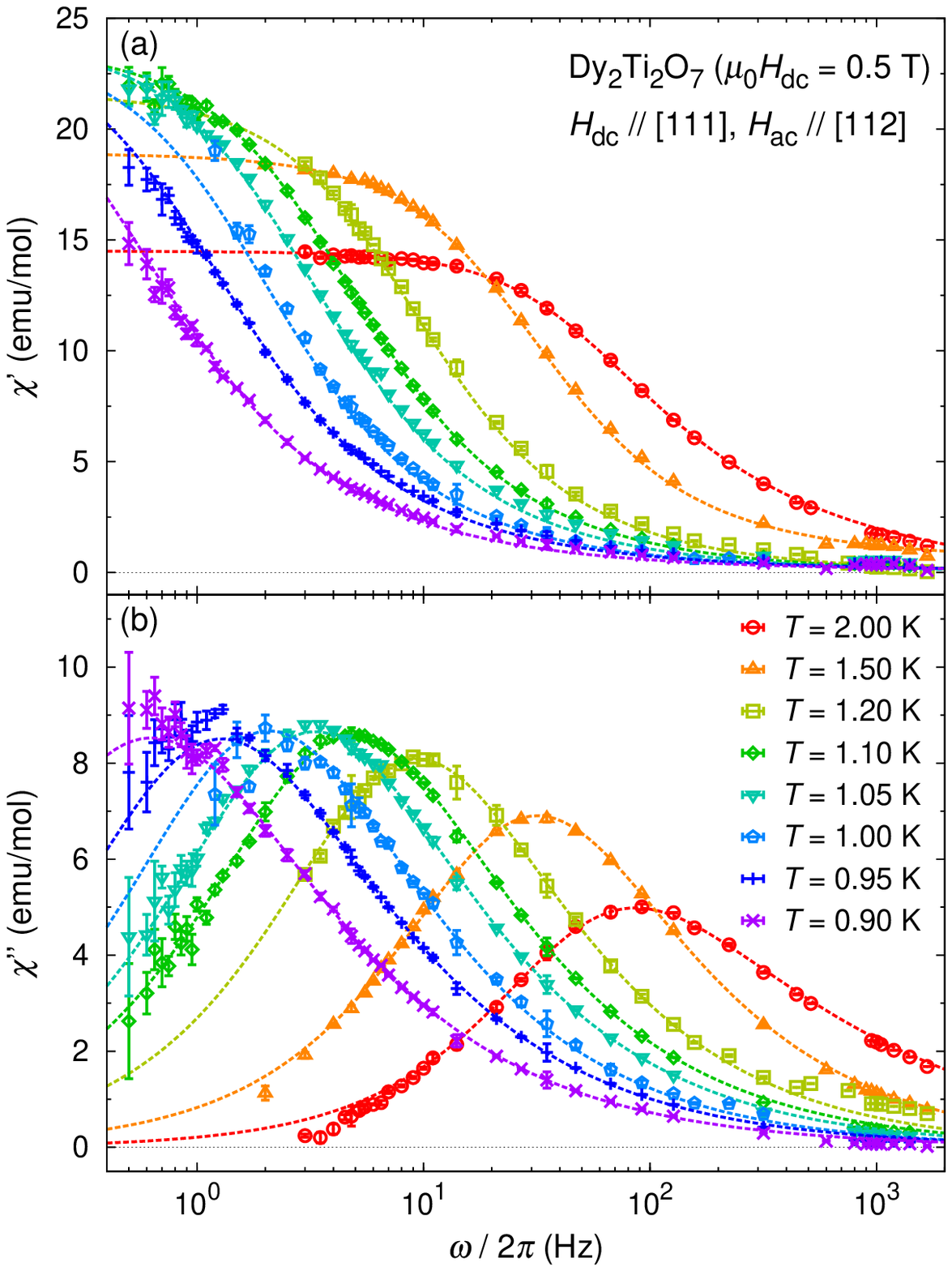}
\caption{
Fitting results (dashed lines) using Eqs.~(S6) and (S7) to 
representative experimental data of (a) $\chi^{\prime}(\omega)$ 
and (b) $\chi^{\prime\prime}(\omega)$, shown in Fig.2 of the main text.
}
\label{fig.S2}
\end{center}
\end{figure}
%%%%%%%%%%%%%%%%%%%%%%%%%%%%%%%%%%%%%%%%%%%%%%%%%%%%%%%%%%%%%%%%%%%%%%%%%%%%%%%%
The dc limit of the real part of the ac magnetic susceptibility $\chi^{\prime}(0)$ was estimated by fitting experimental $\chi(\omega)$ with the following formula known as the Havriliak-Negami relaxation relation~\cite{Havriliak1966_SM,Havriliak1967_SM}:
\begin{align}
&\chi (\omega)         = \chi^{\prime}(\omega)+i\chi^{\prime \prime }(\omega)
                       = \chi_{\mathrm{S}} + \frac{\chi_{\mathrm{T}}-\chi_{\mathrm{S}}}{(1+(i\omega \tau)^{\beta})^{\alpha}}\\
&\chi^{\prime}(\omega) = (\chi_{\mathrm{T}}-\chi_{\mathrm{S}})\Bigl(\frac{\cos(\phi)}{a}\Bigr)^{\alpha} \cos(\alpha \phi) + \chi_{\mathrm{S}}\\
&\chi^{\prime \prime}(\omega) = (\chi_{\mathrm{T}}-\chi_{\mathrm{S}})\Bigl( \frac{\cos(\phi)}{a}\Bigr)^{\alpha} \sin(\alpha \phi)
\label{eq.S4}
\end{align}
where $\chi_{\mathrm{S}} $ and $\chi_{\mathrm{T}}$ are the adiabatic and isothermal
susceptibilities,
$\phi = \arctan (b/a)$, $a=1+(\omega \tau)^\beta \cos(\beta \pi/2)$, 
and $b = (\omega \tau)^\beta \sin(\beta \pi/2)$, respectively. $\alpha$ and $\beta$ are constants between 0 and 1~\cite{K.S.Cole1941_SM,D.W.Davidson1951_SM,D.Gatteschi_SM}. Equation (S5) is an empirical formula where the distribution of relaxation time $\tau$ is considered in the Debye relaxation model in electromagnetism~\cite{Debye_relaxation_paper_SM}:
the case of $\alpha = \beta = 1$ in Eq.~(S5)
corresponds to the Debye relaxation relation with a single dispersion, 
which represents the same formula of the Casimir-du Pr\'{e} relaxation relation~\cite{Casimir1938_SM}.
It is not a precise model for the distribution of $\tau$ but is widely used for the analysis of the frequency dependent susceptibility, since it is written as an analytical form and  qualitatively represents many experimental results~\cite{R.H.Boyd1985_SM,Bauer1996_SM,Richert2000_SM,Asami2002_SM,ToppingJPCM2019_SM}.
As shown in Fig.~S2,
we reasonably fitted the experimental data with Eqs.~(S6) and (S7), 
and obtained $\chi^{\prime}(0)$ as the dc limit of Eq.~(S5) or Eq.~(S6).

\end{document}